\documentstyle[aps,prl]{revtex}

\begin{document}
\pagestyle{empty}
\narrowtext
\twocolumn

\noindent
{\bf Hopf Term for a Two-Dimensional Electron Gas: \\ Comment to Reply
to Comment} \\

In our Comment \cite{VolovikYakovenko} to the Letter
\cite{ApelBychkov} we questioned the derivation  of the Hopf term in
the hydrodynamic action for the Skyrmion dynamics in QHE. Our main
argument was that the description in terms of Euler angles used in
\cite{ApelBychkov} was potentially dangerous since these angles were
ill-defined. The Reply \cite{ApelBychkov2} to our Comment  confirmed
our apprehension. In their Reply \cite{ApelBychkov2} Apel and Bychkov
used their hydrodynamic action to derive the spin of the  Skyrmion
from the adiabatic rotation of the Skyrmion. They obtained the value
of spin $s=(1/2)Q$ where
\begin{equation}
Q={1\over 4\pi} \int d^2r~\{ \partial_x[\cos\bar\theta~
\partial_y\bar\phi] - \partial_y[\cos\bar\theta ~
\partial_x\bar\phi]  \}.
\label{Q}
\end{equation}
They claimed that $Q$ is the topological charge of
Skyrmion. Calculating  $Q$ for the
configuration of the Euler angles
\begin{equation}
\cos\bar\theta(r=\infty)=1~,~
\cos\bar\theta(r=0)=-1~,~
\bar\phi=\arctan (y/x)
\label{configuration}
\end{equation}
which corresponds to the Skyrmion with
the topological charge 1, they obtained $Q=1$ and thus suggested
that their hydrodynamic action does contain the Hopf term responsible
for the spin $1/2$ of the Skyrmion.

Thus the controversy is reduced to the estimation of the integral in
Eq.(1), which is an easy task. Let us check the statement that the
Eq.(1) gives the value $Q=1$ for the configuration in Eq.(2). Though
the angle $\bar\phi$ is ill-defined, its derivatives are well defined.
Thus the integrand in the Eq.(1)  is also well defined  in the whole
2D plane including the origin $r=0$, though it diverges there. That
is why one can easily estimate the integral either by direct
calculations or by converting it to a contour integral.

 Let us calculate the integral in Eq.(1) directly. It can be
expressed as a sum of two terms:
\begin{equation}
Q=Q_1+Q_2~,
\label{Q1+Q2}
\end{equation}
where
\begin{equation}
Q_1={1\over 4\pi} \int d^2r~\{
(\partial_x\cos\bar\theta)(\partial_y\bar\phi)-
(\partial_y\cos\bar\theta)(\partial_x\bar\phi )\},
\label{Q1}
\end{equation}
and
\begin{equation}
Q_2={1\over 4\pi} \int d^2r~\{\cos\bar\theta (  \partial_x\partial_y -
\partial_y\partial_x)\bar\phi \}.
\label{Q2}
\end{equation}

The first integral gives $Q_1=1$. In the second integral one has
\begin{equation}
(\partial_x \partial_y  -\partial_y \partial_x) \bar\phi = 2\pi
\delta({\bf  r}).
\label{delta-function}
\end{equation}
Using the configuration in Eq.(2)
one obtains
$Q_2=-1/2$.  Altogether this gives
\begin{equation}
Q=Q_1+Q_2=1-1/2=1/2 ~~.
\label{Q=1/2}
\end{equation}
instead of $Q=1$.

This mistake in calculation of the integral in Eq. (1) (Eq.(4)  of
\cite{ApelBychkov2})  is rather common, since the
$\delta$-function contribution is often forgotten, and it serves
as a  nice illustration of the potential danger of using  the Euler
angles for the descritption of the Hopf term.  As was observed by
Apel and Bychkov
\cite{ApelBychkov2} the same result
$Q= 1/2$ can be obtained from Eq.(1) by converting the 2D integral to
a contour integral around
$r=\infty$. But they disregarded this  as a wrong way of
calculation of Eq.(1).

Instead Apel and Bychkov suggested an alternative way of the
estimation of the Eq.(1): they introduced the hole in
the 2D space, where the Euler angles have singularity, and considered
the  2D space with removed point.  In this case the volume integral in
Eq.(1) can be
transformed  into the contour integrals at infinity and around the
infinitely thin cylindrical surface embracing the hole. Such
description in terms of the holes gives the correct value for the
topological invariant $Q=1/2 + 1/2=1$. But if one introduces holes
for the calculation of topological invariant $Q$, one should also
introduce the holes in the 3D space-time (the space-time with removed
lines) for the calculation of the Hopf invariant
in action: one cannot use different approaches for the calculation of
$Q$ and of Hopf invariant! In this case the problem becomes
complicated, since one must estimate the contribution of these
dynamical holes into the hydrodynamic action. Apparently this was
not made in \cite{ApelBychkov}: the hole
contribution has not been considered there.

Conclusion:

(i) If the  holes in 2D space are introduced, the calculations of the
Hopf term, made by Apel and Bychkov  \cite{ApelBychkov}, should be
reconsidered.

(ii) If the holes are not introduced, then the Eq.(1)  should
be valid in the whole 2D space and thus the Hopf-like action
calculated in \cite{ApelBychkov} gives the wrong result
$Q=1/2$ for  Skyrmion with topological charge $Q=1$.

In both cases the  derivation of the Hopf term in
\cite{ApelBychkov} remains questionable.  \\

\noindent
G.~E.~Volovik

Helsinki University of Technology, Low Temperature Laboratory,
P. O. Box 2200, FIN-02015 HUT, Finland

Landau Institute for Theoretical Physics, 117334 Moscow, Russia\\
\\
November 10, 1997 

\end{document}